 \documentclass[final,3p,times,twocolumn]{elsarticle}

\usepackage{graphicx}
\usepackage{latexsym}
\usepackage{amsfonts}
\usepackage{amssymb}
\usepackage{amsmath}
\usepackage{bm}
\usepackage{mathrsfs}
\usepackage{slashed}
\usepackage{ascmac}
\usepackage{comment}

\usepackage{dcolumn}

\allowdisplaybreaks[3] 

\usepackage[utf8]{inputenc} 

\renewcommand{\and}{{\rm and}}

\newcommand{\cout}[1]{ \if 0 {#1} \fi }

\renewcommand{\=}{&=&}

\newcommand{\bq}{{\bm{q}}}

\newcommand{\bB}{{\bm B}}

\newcommand{\para}{ \parallel}

\usepackage{color} 
\usepackage[normalem]{ulem}  

\begin{document}

\begin{frontmatter}



\title{Physics of strong electromagnetic fields in relativistic heavy-ion collisions}


\author{Koichi Hattori} 

\affiliation{organization={Zhejiang Institute of Modern Physics, Department of Physics, Zhejiang University},
            addressline={}, 
            city={Hangzhou},
            postcode={310027}, 
            state={Zhejiang},
            country={China}}
            
\affiliation{organization={Research Center for Nuclear Physics, Osaka University},
            addressline={10-1 Mihogaoka}, 
            city={Ibaraki},
            postcode={567-0047}, 
            state={Osaka},
            country={Japan}} 

\begin{abstract}
I discuss several roles of the strong electromagnetic fields created by relativistic heavy-ion collisions. These phenomena call for theoretical and experimental developments to understand dynamics of quark-gluon plasma (QGP) 
as well as purely electromagnetic processes in the ultraperipheral collisions. 
\end{abstract}
%


\begin{keyword} Relativistic heavy-ion collisions, 
Quark-gluon plasma, Strong magnetic field



\end{keyword}

\end{frontmatter}








%
%
%
%
%
%

\section{Introduction}


Relativistic heavy-ion collisions at RHIC and the LHC create the extreme state of matter called quark-gluon plasma (QGP) that is a deconfined state of quarks and gluons. One remarkable feature accompanying these collisions is the generation of extremely strong electromagnetic fields, with magnetic field strengths reaching $10^{18} - 10^{19} $ 
Gauss in non-central collisions (see Refs.~\cite{Deng:2012pc, Hattori:2016emy} and references therein). These fields exist transiently but are expected to be strong enough to influence the early-time dynamics of the system and various measurable observables, thereby opening a new frontier in the study of QGP under extreme conditions and an interdisciplinary study of quantum systems under strong fields \cite{Hattori:2023egw}.


The presence of strong electromagnetic fields introduces anisotropies and novel quantum effects into the dynamics of QGP. In particular, electromagnetic fields modify the behavior of hard probes, such as photons, dileptons, and heavy quarks, through vacuum polarization, medium-induced dispersion, and diffusion dynamics. Real and virtual photons acquire polarization-dependent refractive indices, leading to phenomena called the vacuum birefringence and dichroism \cite{Hattori:2012je,Hattori:2012ny, Hattori:2020htm}, which have been extensively studied in high-intensity laser physics \cite{Fedotov:2022ely} and astrophysics \cite{Enoto:2019vcg}, and are revisited in the context of relativistic heavy-ion collisions. Similarly, heavy quarks interact with the magnetic field through modified diffusion dynamics as well as 
a direct coupling via the Lorentz force. 


In the soft sector, the magnetic field influences the hydrodynamic evolution of the medium. Recent progress in relativistic magnetohydrodynamics (MHD) has enabled the study of collective modes in anisotropic and conducting fluids with dynamical electromagnetic fields. These advances have clarified how magnetic fields affect transport coefficients and wave propagation \cite{Fang:2024skm}, resolving previous inconsistencies in the mode propagating transverse to the magnetic field. Moreover, extensions to include spin degrees of freedom, spin magnetohydrodynamics have shown that the spin-orbit coupling and angular momentum conservation laws give rise to new transport structures, including not only rotational viscosities and gapped spin modes \cite{Hattori:2019lfp} but also a cross term between the symmetric and antisymmetric parts of the energy-momentum tensor \cite{Fang:2024sym, Fang:2024hxa}.

An important class of phenomena arising in the hydrodynamic regime is anomalous transport phenomena (see, e.g., Ref.~\cite{Hattori:2016emy} for a review), where quantum anomalies induce currents or charge separations. The magnetovortical matter under the coexistent magnetic field and angular momentum, in particular, leads to charge redistribution and polarization phenomena \cite{Hattori:2016njk}, via coupling to both spin and orbital angular momentum. Recent developments have highlighted the importance of properly accounting for the orbital angular momentum associated with the cyclotron motion, which inverts the direction of induced charges and currents compared to earlier predictions based solely on spin contributions \cite{Fukushima:2024tkz}.

This contribution provides an overview of these developments, with an emphasis on both theoretical formulations and connections to experimental observables. By bridging the dynamics of strong fields across hard probes, collective flow, and quantum anomalies, this research area links heavy-ion physics with broader domains such as condensed matter systems, high-intensity laser physics, and astrophysics. Understanding the interplay between QCD matter and electromagnetic fields not only deepens our knowledge of the QGP but also enriches our perspective on fundamental interactions in extreme environments.

\section{Hard probes}

Hard probes, such as photons, dileptons, and heavy quarks, offer insight into both vacuum and medium effects. The interplay between strong electromagnetic fields and QCD matter is crucial particularly in probing the early stages of the collisions. 


\subsection{Photons and dileptons}

Strong magnetic fields modify photon properties through 
the coupling to charged fermion and antifermion pairs 
in the vacuum polarization diagram. 
Photon dispersion relations depend on the polarization modes 
and a propagation direction when response of the fermions to 
propagating electromagnetic fields is anisotropic. 
This is indeed the case for charged particles in a magnetic field 
due to the Lorentz force. 
Polarization-dependent refractive indices are called 
the {\it vacuum birefringence} after a similar optical phenomenon 
observed in materials with anisotropic lattice structures \cite{Toll:1952rq, Adler:1971wn}. 
Moreover, real photons, as well as virtual photons, can decay into 
fermion and antifermion pairs in a strong magnetic field; 
The energy and momentum conservations are both satisfied 
when the fermion energy levels are quantized in the Landau levels \cite{Hattori:2012je, Hattori:2020htm}. 
This is called the {\it vacuum dichroism} again after a similar optical phenomenon. 
The vacuum brefringence and dichroism are quantified by 
the real and imaginary parts of the vacuum polarization tensor, respectively,  
that are related with each other via the Kramers-Kronig relation  \cite{Hattori:2023egw}, and thus are both sides of the same coin. 
Experimental observation of these phenomena has been pursued 
for a long time with high-intensity laser field \cite{Fedotov:2022ely} and radiation from 
magnetospheres of neutron stars and magnetars \cite{Enoto:2019vcg}. 
Ultraperipheral collision events provide another opportunity 
with the ever strongest magnetic field \cite{STAR:2019wlg}. 


The photon dispersion relation is described by the Maxwell equation 
\begin{eqnarray}
[ q^2 g^{\mu\nu} - q^\mu q^\nu -  \Pi^{\mu\nu} ] A_\nu (q) =0
\end{eqnarray}
with the vacuum polarization tensor $  \Pi^{\mu\nu} $ 
in a constant external magnetic field. 
The polarization dependence and anisotropy of photon dispersion relations 
can be understood from the anistropic tensor structures in the vacuum polarization tensor as follows. 
Under a strong magnetic field in vacuum, the vacuum polarization tensor 
has three independent structures \cite{Dittrich:2000zu, Hattori:2012je}
\begin{eqnarray}
\Pi^{\mu\nu} = -[ \chi_0 P_0^{\mu\nu}
+ \chi_1 P_\para^{\mu\nu} + \chi_2 P_\perp^{\mu\nu}] \, ,
\end{eqnarray}
which are constrained by the Ward identity $ q_\mu \Pi^{\mu\nu} =0$ 
with a photon momentum $ q^\mu$. 
While $ P_0^{\mu\nu}$ is the familiar gauge-invariant structure, 
a strong magnetic field gives rise to two additional structures 
$ P_{\para/\perp}^{\mu\nu} = q_{\para/\perp}^2 g_{\para/\perp}^{\mu\nu} - q_{\para/\perp}^\mu q_{\para/\perp}^\nu$, where 
$ \eta^{\mu\nu}_\para = {\rm diag} (1,0,0,-1)$, 
$\eta^{\mu\nu}_\perp = {\rm diag} (0,-1,-1,0)$, 
and $q^\mu_{\para/\perp} = g_{\para/\perp}^{\mu\nu}q_\nu$. 
Plugging the above tensor decomposition into the Maxwell equation 
and fixing the gauge redundancy appropriately, 
we readily find the polarization-dependent refractive indices 
$n = |\bq|/q^0 $ as  
\begin{subequations}
\label{eq:}
\begin{eqnarray} 
n_\para^2 \= 
\frac{1+\chi_0+\chi_1}{1+\chi_0+\chi_1\cos^2\theta}
\, ,   \\
n_\perp^2 \=
\frac{1+\chi_0}{1+\chi_0+\chi_2\sin^2\theta}
\, ,
\end{eqnarray}
\end{subequations}
where $ \theta$ is a relative angle between the photon momentum 
and the magnetic field direction. 
Both refractive indices reduce to unity when $\theta=0 $ 
in accordance with the residual Lorentz-boost invariance along a constant magnetic field 
and reduce to unity in the vanishing field limit $B\to0 $ 
where two of the coefficients $ \chi_{1,2}$ vanish. 

The Landau-level representation of the coefficients $ \chi_{0,1,2}$ 
were shown in Refs.~\cite{Hattori:2012je}. 
Extensions to finite temperature and density are investigated recently  \cite{Hattori:2022uzp, Hattori:2022wao, Fukushima:2024ete}. 
Photon and dilepton spectrum are also discussed in e.g., Refs.~\cite{Hattori:2012ny, Ishikawa:2013fxa, Hattori:2020htm, Wang:2020dsr, Wang:2024gnh} 
and are recently combined with hydrodynamic simulations \cite{Kimura:2024gao}. 
In addition to the polarization and angle dependences of the spectrum 
discussed in these works, dominance of the muon pair yield over 
the electron pair yield signals the dilepton production in a strong magnetic field in analogy with the ``helicity suppression'' 
mechanism known in the charged pion decay \cite{Hattori:2020htm}.

\subsection{Heavy quarks}

Dynamics of heavy quarks have been discussed as probes of QGP properties. 
Effects of the magnetic fields on heavy-quark dynamics 
should be discussed in each stage of the time evolution. 
The first stage is the heavy-quark production discussed in Ref.~\cite{Chen:2024lmp} and then competing effects of the Lorentz force 
and the Faraday effect may occur in a time-dependent magnetic field 
as discussed in Refs.~\cite{Das:2016cwd, Sun:2023adv}. 
After the formation of QGP, an anisotropic diffusion constant 
for the Brownian motion of heavy quarks 
becomes important \cite{Fukushima:2015wck}. 
These effects modify the open heavy-quark spectrum in the final state \cite{STAR:2019clv}.


\section{Low-energy dynamics in hydrodynamic regime}

Hydrodynamics describes slow dynamics of conserved charges. 
Time evolution of QGP has been modeled with hydrodynamics. 
Magnetohydrodynamics (MHD) includes the magnetic flux conservation 
and describes intertwined dynamics of 
QGP and a dynamical electromagnetic field (see Ref.~\cite{Hattori:2022hyo} for a recent review). 
Effects of quantum anomaly have been also intensively studied 
in the two decades (see review articles \cite{Hattori:2016emy, Hattori:2023egw} and references therein).

\subsection{Relativistic magnetohydrodynamics}

Relativistic MHD was reformulated based on 
the magnetic-flux conservation \cite{Grozdanov:2016tdf,Hattori:2017usa, Hongo:2020qpv,Hattori:2022hyo}
\begin{eqnarray}
\partial_\mu T^{\mu\nu}_{\rm total} =0 \, , 
\quad 
\partial_\mu \tilde F^{\mu\nu} = 0 \, ,
\end{eqnarray}
where $ T^{\mu\nu}_{\rm total} $ and $ \tilde F^{\mu\nu}$ are the total energy-momentum tensor including the electromagnetic contribution and 
the dual field strength tensor, respectively. 
A preferred orientation provided by a magnetic field 
gives rise to anisotropic transport coefficients. 
Recently, a complete set of linear waves are, for the first time, 
obtained for a general propagation direction 
with all the anistoropic first-order transport coefficients \cite{Fang:2024skm}. 
An inconsistency between the preceding results at the transverse angle 
has been understood based on this general solution. 
In fact, the angle dependence has a singularity near the transverse direction 
due to the anistoropy. 
This is potentially a general issue in anistoropic systems. 

Numerical simulations with a dynamical magnetic field were implemented recently by two groups~\cite{Nakamura:2022ssn, Mayer:2024dze, Mayer:2024kkv,Benoit:2025amn}. 
These studies are indispensable for understanding the charge-dependent flows 
observed at the LHC and RHIC \cite{ALICE:2019sgg, STAR:2023jdd}. 
In applications of MHD, it is important to 
know the transport coefficients in a magnetic field. 
In the strong-field regime, the leading-order scattering channel 
is provided by a 1-to-2 real-photon radiation from a fermion \cite{Hattori:2016lqx,Hattori:2016cnt, Hattori:2017qih, Fukushima:2017lvb, Ghosh:2024hbf, Ghosh:2024fkg}. 
This is a cross channel of the photon decay discussed above, 
and contrasts to the conventional 2-to-2 scattering channels in the absence of a magnetic field \cite{Arnold:2000dr} or in a weak magnetic field \cite{Li:2017tgi}. 
See more relevant studies in, e.g., Refs.~\cite{Mohanty:2018eja,Das:2019ppb,Dey:2019axu,Das:2019pqd,Chen:2019usj,Dash:2020vxk,Rath:2020beo,Panda:2020zhr,Panda:2021pvq,Rath:2021ryd,Rath:2022oum, Ghosh:2022xtv}.

Including spin degrees of freedom leads to formulation of spin MHD \cite{Fang:2024skm}. In the presence of a magnetic field counted as a zeroth-order quantity in gradient, there appears a new cross term between the symmetric and antisymmetric parts of the energy-momentum tensor. 
This effect may be similar to the Magnus effect where a force is exerted on a spinning object in perpendicular to the spin axis under a hydrodynamic flow. 
Understanding spin transport is important for the experimental observation of global spin polarization at RHIC \cite{STAR:2017ckg, STAR:2023nvo} or lower-energy collisions in future.


%
%
%

\subsection{Magnetovortical matter}

Interplay between effects of a magnetic field and an angular velocity 
induces a charge density \cite{Hattori:2016njk}. 
This effect is identified as a family of ``anomalous transport phenomena''  induced by quantum anomaly (see Ref.~\cite{Yamamoto:2021gts} 
and references in Ref.~\cite{Fukushima:2024tkz}). 
To understand this effect intuitively, notice that 
spin is coupled with both a magnetic field and an angular velocity. 
However, the former electromagnetic coupling depends on signs of electric charges, while the latter mechanical one is blind to electric charges. 
When we start rotating a system adiabatically under a strong magnetic field, 
particles and antiparticles acquire opposite energy shifts 
due to the opposite spin polarization in a strong magnetic field. 
The induced energy difference can be interpreted as an effective chemical potential.

This effect has been reexamined recently in Ref.~\cite{Fukushima:2024tkz}. 
One crucial contribution overlooked in Ref.~\cite{Hattori:2016njk} is 
the orbital angular momentum associated with the Landau levels. 
As clear in classical mechanics, each cyclotron orbit 
possesses an orbital angular momentum, albeit quantized, when 
its expectation value is computed with the Landau-level basis \cite{Hattori:2023egw}. 
The orbital contribution is diamagnetic, 
while the spin contribution is paramagnetic. 
The orbital contribution overwhelms the spin contribution (because the magnitude of mechanical coupling to spin does not have the $ g$ factor). 
As a result, the sign of induced electric charge is inverted as compared to 
the previous results as 
\begin{eqnarray}
j^0 = C_A \big( \frac{1}{2} -1 \big) {\bm \omega} \cdot \bB
= -   \frac{C_A}{2} {\bm \omega} \cdot \bB  \, ,
\end{eqnarray}
with the anomaly coefficient for Dirac fermions $ C_A = e^2/(2\pi^2)$ 
and the vorticity $ {\bm \omega} $. 
The positive and negative terms between the parentheses 
are the spin and orbital contributions, respectively. 
A gauge-invariant formulation of thermodynamic partition function 
applied to a 
steady state leads to the prediction of the sign inversion \cite{Fukushima:2024tkz}. 
It should be emphasized that, for a rotating matter to reach 
a steady state under a magnetic field, 
the radial drift force {\it induced by rotation} in a magnetic field 
should be offset by a radial electric field. 
Namely, in the steady state, there exist both electric and magnetic fields 
in the inertial frame for the balance of force in the radial direction; 
Equivalently, there is no electric field in the comoving frame. 



\section{Summary}

In this contribution, I have surveyed the facets of strong-field physics in relativistic heavy-ion collisions. The strong electromagnetic fields created in the early stages of non-central collisions induce a wide spectrum of physical phenomena across both hard and soft probes. Their coupling to photons and heavy quarks alters fundamental properties such as the dispersion relations, decay rates, and diffusion constants. On the other hand, in the hydrodynamic regime, the presence of magnetic and vortical fields enriches the spectrum of collective excitations, particularly when spin degrees of freedom are included.

The interplay of electromagnetic fields with quantum anomalies has also opened new directions in understanding charge separation, polarization, and transport phenomena in not only QGP but also condensed matter physics with the Dirac and Weyl semimetals. 
Very recently, revisiting the interplay between a magnetic field and an angular velocity uncovered the crucial contribution of the orbital angular momentum that has been overlooked for almost a decade. 
The orbital contribution inverts the sing of the overall effect 
in the strong magnetic field.


In perspective, cooperative and continued advances in numerical simulations, analytic formulations, and experimental observations are essential for fully realizing the potential of strong-field physics in heavy-ion collisions. This research area also provides opportunities for interdisciplinary engagement with research fields such as high-intensity laser physics, astrophysics, and condensed matter physics which offer analogous physical platforms. Ultimately, by studying the response of QCD matter to strong fields, we not only gain a deeper understanding of QGP but also a unified view of quantum many-body systems under extreme conditions.

\section*{Acknowledgments} 
This work is supported in part by JSPS KAKENHI under grant No.~23K22487. 

\bibliographystyle{elsarticle-num} 
\bibliography{ATHIC2025-bib}

\end{document}